\documentclass[aps,showpacs,superscriptaddress,citeautoscript,preprint,prb,floatfix]{revtex4-1}

\usepackage{bm}
\usepackage{graphicx}
\usepackage{amsmath}
\usepackage{amsfonts}
\usepackage{amssymb}
\usepackage{siunitx}
\usepackage{color}

\def \Vcal {\mathcal{V}}

\def \alc {\alpha_{\mathrm{c}}}

\begin{document}

\title{Spin-polarized electron transmission in DNA-like systems}

\author{Miguel A. Sierra}

\affiliation{Institute for Cross-Disciplinary Physics and Complex Systems IFISC (UIB-CSIC), E-07122 Palma de Mallorca, Spain}

\affiliation{Institut f{\"u}r Theoretische Physik (TP4) and W{\"u}rzburg-Dresden Cluster of Excellence ct.qmat, Universit{\"a}t W{\"u}rzburg,97074 W{\"u}rzburg, Germany}

\author{David S\'{a}nchez}

\affiliation{Institute for Cross-Disciplinary Physics and Complex Systems IFISC (UIB-CSIC), E-07122 Palma de Mallorca, Spain}

\author{Rafael Gutierrez}

\affiliation{Institute for Materials Science and Max Bergmann Center of Biomaterials, TU Dresden, 01062, Dresden, Germany}

\author{Gianaurelio Cuniberti}

\affiliation{Institute for Materials Science and Max Bergmann Center of Biomaterials, TU Dresden, 01062, Dresden, Germany}

\affiliation{Dresden Center for Computational Materials Science, TU Dresden, 01062 Dresden, Germany}

\affiliation{Center for Advancing Electronics Dresden, TU Dresden, 01062 Dresden, Germany}

\author{Francisco Dom\'{i}nguez-Adame}

\affiliation{GISC, Departamento de F\'{\i}sica de Materiales, Universidad Complutense, E--28040 Madrid, Spain}

\author{Elena D\'{\i}az}

\affiliation{GISC, Departamento de F\'{\i}sica de Materiales, Universidad Complutense, E--28040 Madrid, Spain}

\begin{abstract}

The helical distribution of the electronic density in  chiral molecules, such as DNA and bacteriorhodopsin, has been suggested to induce a spin-orbit coupling interaction  that may lead to the so-called chirality-induced spin selectivity (CISS) effect. Key ingredients for the theoretical modelling are, in this context, the helically shaped potential of the molecule and, concomitantly, a Rashba-like spin-orbit coupling due to the appearance of a magnetic field in the electron reference frame. Symmetries of these models clearly play a crucial role in explaining the observed effect, but a thorough analysis has been largely ignored in the literature. In this work, we present a study of these symmetries and how they can be exploited to enhance chiral-induced spin selectivity in helical molecular systems.

\end{abstract}

\begin{flushright} 
\emph{Biomolecules} \textbf{10}, 49 (2020)
\end{flushright}

\maketitle

\section{Introduction}

The discovery of the spin polarization capability of helical molecules a few years ago~\cite{Gohler2011,doi:10.1021/nl2021637} has demonstrated an intriguing novel physical phenomenon, which has been called chirality-induced spin selectivity (CISS). Although work on spin-dependent effects in chiral systems can be traced back to earlier  work by Kessler and others (see Ref.~\cite{kessler}), it was not until 2011 that the previously mentioned two works clearly showed strong spin polarization effects in chiral molecules (DNA in this case) using two different experimental approaches: photoemission experiments~\cite{Gohler2011} and AFM-based electrical transport setups~\cite{doi:10.1021/nl2021637}. The CISS effect is very striking since it does not require the presence of any magnetic centers or strong spin-orbit coupling (SOC) effects in the systems where it has been observed~\cite{ADMA:ADMA201504725,C6CS00369A,Mishra14872,doi:10.1063/1.4966237,SMLL:SMLL201602519,Kumar2017,doi:10.1021/acsnano.5b00832,doi:10.1021/jp509974z,doi:10.1146/annurev-physchem-040214-121554,doi:10.1021/jacs.6b10538,doi:10.1021/jacs.6b12971,BenDor2017,doi:10.1021/acs.jpclett.8b00208,doi:10.1021/jacs.8b08421}. Its universality in molecular systems displaying a helical shape has put forward the hypothesis that a very close relationship must exist between helical symmetry and spin selectivity. Many theoretical works have been devoted up to now to scrutinize the CISS effect, largely based on spin-dependent transport calculations using scattering matrix or Green's function techniques~\cite{Yeganeh09,Medina12,Gutierrez12,Gutierrez13,doi:10.1021/acs.jpclett.8b02196,Eremenko13,Medina15,Caetano16,Guo12,Guo14a,Diaz17bis,Matityahu16,doi:10.1063/1.4820907,doi:10.1021/acs.nanolett.9b01707,doi:10.1021/acs.jpcc.9b07764,Michaeli15,doi:10.1021/acs.jpclett.9b02929}. Also few first-principle calculations have been presented~\cite{doi:10.1021/acs.jpclett.8b02360,carmen}, further supporting the relation to the helical symmetry. Still, the debate has remained open and there is not a common theoretical framework explaining the CISS effect yet. Most of the previously cited theoretical investigations based on model Hamiltonian approaches assume the presence of some type of generic SOC in the molecular systems and, based on it, proceed to discuss its consequences for the CISS effect. A major issue has been whether spin polarization can be found in a two-terminal setup without invoking dephasing~\cite{Guo12} or non-unitarity effects~\cite{Matityahu16}. Some indications that these factors may not be required were given in Ref.~\cite{Gutierrez13} and more recently in Ref.~\cite{doi:10.1021/acs.jpclett.9b02929}, but the discussion remains also open on this point.

In this paper, we consider a very generic model previously used to describe the CISS effect~\cite{Gutierrez13,Guo12}, which consists of two inter-connected tight-binding chains, mimicking two interacting helices, including spin-orbit interaction and attached to two fermionic reservoirs playing the role of current terminals. We discuss in detail the general symmetries of the model ---an issue not addressed previously--- and show the parameter ranges where spin polarization can be important. It is interesting to note that our model does not assume any source of dissipation in the molecule. All dissipation takes place deep in the reservoirs, which are assumed to be in local thermodynamic equilibrium with well defined electrochemical potential and temperature. In the next section, we introduce the model and discuss the system energy spectrum by analytically diagonalizing the Hamiltonian. Using non-equilibrium Green's function techniques, the spin-dependent electrical current is derived from the transmission probability for electrons traversing the two-terminal setup. Thus, we can define an energy-dependent spin polarization function. 
Special attention is devoted to the general symmetry properties of the spin polarization. Furthermore, we find that spin-flip processes do not contribute to the computed spin polarization, in perfect agreement both with first-principle calculations~\cite{doi:10.1021/acs.jpclett.8b02360} and with the analysis of photoemission experiments~\cite{C8CP07257D}.

\section{Results}
\subsection{Tight-binding model with generalized Rashba interaction}
\label{TB}

Our system Hamiltonian describes two different helical strands $\mu=\{A,B\}$ with sites $n=\{1,2, \ldots, N\}$ that can be occupied with electrons of spin $\sigma = \{\uparrow, \downarrow \}$. Including intrastrand ($t_\mu$) and interstrand hoppings ($\alpha_c$) as well as a non-zero spin-orbit coupling, we split the system Hamiltonian as 
$\mathcal{H}_{\rm} = 
\mathcal{H}_{\rm t} + \mathcal{H}_{\rm soc} + \mathcal{H}_{\rm th}$, with
\begin{subequations}\label{Ham}
\begin{eqnarray}
\mathcal{H}_{\rm t} &=& - \sum_{\mu n} t_\mu\left[ d^\dagger_{\mu n} d_{\mu n+1}^{} +d^\dagger_{\mu n+1} d_{\mu n}^{}\right]\ , \\
\mathcal{H}_{\rm soc} &=& i \sum_{\mu n } \lambda_\mu \left[ d^\dagger_{\mu n+1} (\hat{\sigma}^{\perp,\mu n} +\hat{\sigma}^{\perp,\mu n+1} ) d_{\mu n}^{} -   d^\dagger_{\mu n} (\hat{\sigma}^{\perp,\mu n}+\hat{\sigma}^{\perp,\mu n+1})  d_{\mu n+1}^{}\right]\ ,\\
\mathcal{H}_{\rm th} &=& \alpha_{\rm c} \sum_{n} \left(d^\dagger_{A n} d_{B n}^{}+d^\dagger_{B n} d_{A n}^{}\right)\ ,
\end{eqnarray}
\end{subequations}
where the creation and annihilation electron operators are $d^\dagger_{\mu n}=(d^\dagger_{\mu n \uparrow},d^\dagger_{\mu n \downarrow})$ and $d_{\mu n}=(d_{\mu n \uparrow},d_{\mu n \downarrow})^\intercal$, respectively (the superscript $\intercal$ stands for transpose). The energy of the molecular orbitals in the two strands has been set to zero for simplicity. However, the interstrand coupling $\alpha_{\rm c}$ needs to be nonzero to achieve non-vanishing spin polarization, as demonstrated in Refs.~\cite{Gutierrez13} and~\cite{Guo12}. The SOC of the  Hamiltonian with strength $\lambda_\mu$ depends on the spin operator perpendicular to the helical path which is written in terms of the Pauli matrices as follows
\begin{eqnarray}\label{Eq:SigmaPerp}
\hat{\sigma}^{\perp,\mu n} =\hat{\sigma}^x \sin{\varphi_{\mu n}} \sin{\theta} - \hat{\sigma}^y \cos{\varphi_{\mu n}} \sin{\theta} + \hat{\sigma}^z \cos{\theta}\ ,
\end{eqnarray}
where $\theta$ is the helix angle, and $\varphi_{A n} = n\varphi$ and $\varphi_{B n} = n\varphi+\pi$ are the azymuthal angles in the two helical strands for the B-form of DNA~\cite{Guo12}. 

\subsection{Band structure of a molecule with site-independent SOC}
\label{Bands}

As described in Eq.~\eqref{Ham}, the molecular SOC is site-dependent. However, when one assumes a site-independent SOC ($\varphi_{An} = \varphi_A$ and $\varphi_{Bn} = \varphi_B = \varphi_A + \pi$), this model still allows for a finite SOC to exist. Within this approximation, the electron band structure can be easily calculated in the standard way by performing a Fourier transform. The matrix Hamiltonian in momentum space is then
\begin{eqnarray}
\mathcal{H}(k) &=& \begin{pmatrix}
E_{A+}(k) & 0 & \alc \cos{\theta} & \alc \sin{\theta}\\
0 & E_{A-}(k) &  \alc \sin{\theta} &  -\alc \cos{\theta} \\
\alc \cos{\theta} &  \alc \sin{\theta} & E_{B+}(k) & 0\\
\alc \sin{\theta} & -\alc \cos{\theta}& 0 & E_{B-}(k)
\label{HPi}
\end{pmatrix} \ ,
\end{eqnarray}
where $E_{\mu S} = 4S\lambda_\mu \sin{k} - 2t_\mu \cos{k}$ with $S=\pm$ and the wavenumber $k$ lies between $-\pi$ and $\pi$ for the first Brillouin zone.

\begin{figure}[ht]
\centering
\includegraphics[scale=0.8]{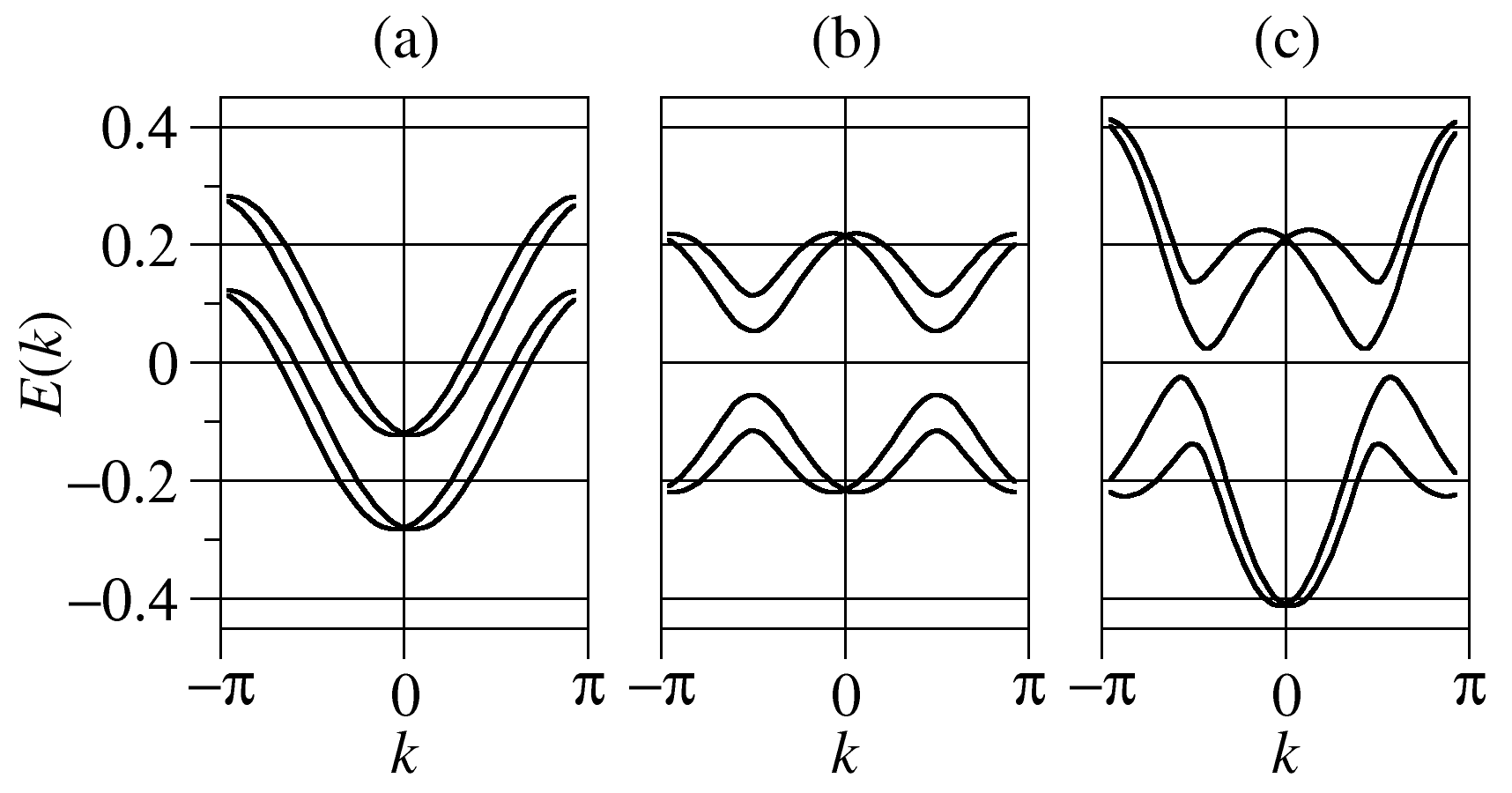}
\caption{Energy dispersion relation of the helical system for $\alpha_c = -0.08$ and $\theta = \pi/6$ in (\textbf{a})~the symmetric configuration with parameters $t_A=t_B=0.1$ and $\lambda_A = \lambda_B = 0.01$,  (\textbf{b})~the quasi-symmetric configuration with parameters $t_A=-t_B=0.1$ and $\lambda_A =\lambda_B = 0.01$, and (\textbf{c})~the asymmetric configuration with parameters $t_A=-2t_B=0.2$ and $2\lambda_A = \lambda_B = 0.02$.}
\label{Fig:0}
\end{figure}

In figure~\ref{Fig:0} we plot the energy bands calculated from the diagonalization of equation~\eqref{HPi} for three different situations satisfying the requirement $\lambda_{\mu} < |\alc|$ $\sim$ $|t_{\mu}|$ that will be analyzed hereafter. Notice that the condition $\Delta \varphi = \varphi_A-\varphi_B = \pi$ accounts for a physical scenario where electrons propagating along the two helices, A and B, perform a precession around opposite magnetic fields within the $XY$ plane.  The first case under consideration will be referred to as the symmetric one since $t_A=t_B$ and $\lambda_A =\lambda_B$. Furthermore, we will also be dealing with asymmetric cases such as $t_A=-xt_B=$ and $x\lambda_A = \lambda_B$, $x$ being a scaling factor. The latter has been demonstrated as the most favorable case to obtain an enhanced spin polarization~\cite{Gutierrez13, Guo14a}.

In the symmetric case [see figure~\ref{Fig:0}(a)], the tight-binding bands undergo a horizontal splitting due to the SOC field. This finding is consistent with the shifts observed in Rashba quantum wires with parabolic dispersions~\cite{Moroz99}. Two separated groups of bands are obtained due to the interstrand hopping term. The asymmetric cases (see figures~\ref{Fig:0}(b) and~(c)) are more involved because the nature of the eigenstates (bonding or antibonding) differ in the two strands. Noticeably, a small gap opens for energies around the reference energy. In each case, we recover the spin degeneracy for zero wavenumber. Our results are qualitative since we neglected the site dependence in the SOC potential. However, we do not expect significant departures in the full dependence case.

\subsection{Transmission and spin polarization}
\label{TransSP}

We now discuss charge transport when the molecule is tunnel coupled to two non-magnetic terminals attached at the edges of the helical molecule. We model such connection by adding the following two terms to the system Hamiltonian (\ref{Ham})
\begin{subequations}
\begin{eqnarray}
\mathcal{H}_{\rm lead} &=& \sum_{\alpha k} \varepsilon_{\alpha k} C^\dagger_{\alpha k} C_{\alpha k}\ ,\\
\mathcal{H}_{\rm tun} 	&=& 	\sum_{\alpha k \sigma\mu } \left[\Vcal_{\alpha k, \mu n} C^\dagger_{\alpha k } d_{\mu n } +\Vcal_{\alpha k, \mu n}^*  d_{\mu n }^\dagger C_{\alpha k }\right] \ ,
\end{eqnarray}
\end{subequations}
where the creation and annihilation electron operators at the leads are $C^\dagger_{\alpha k}=(C^\dagger_{\alpha k \uparrow},C^\dagger_{\alpha k \downarrow})$ and $C_{\alpha k}=(C_{\alpha k \uparrow},C_{\alpha k \downarrow})^\intercal$, respectively. Here, the tunneling amplitude $\Vcal_{\alpha k, \mu n}$ is nonzero only when $\{\alpha=L,  n=1\}$ and $\{\alpha = R, n=N\}$. Using well-known techniques of non-equilibrium transport calculations~\cite{JauhoBook}, after a lengthy but straightforward algebra we find the electronic transmission probability in terms of advanced and retarded Green's functions:
\begin{equation}
\mathcal{T}(\omega) = \sum_{ \mu \nu \rho\tau s}\Gamma^{R}_{ \nu 1 \sigma,\mu 1 \sigma}(\omega) G^r_{\mu 1 \sigma,\rho N s}(\omega) \Gamma^{L}_{ \rho N s,\tau N s}(\omega) G^a_{\tau N s,\nu 1 \sigma}(\omega)\ .
\end{equation}

Our objective is to calculate the transmission probability per spin and strand, which we define as
\begin{eqnarray}\label{Eq:Transmission}
\mathcal{T}_{\mu \sigma,\nu s}(\omega) =  \Gamma^R_{\mu 1 \sigma, \mu 1 \sigma} G^r_{\mu 1 \sigma, \nu N s} \Gamma^L_{\nu N s, \nu N s} G^a_{\nu N s, \mu 1 \sigma} \ ,
\end{eqnarray}
where the retarded and advanced Green's functions are evaluated from the expression $G^{r,a}(\omega) = [\omega-\mathcal{H} - \Sigma^{r,a}]^{-1}$. This is exact for independent electrons as the ones considered in this work. $\Sigma^{r,a}$ are the retarded and advanced electron self-energies due to tunneling. Their matrix elements are the broadening coefficients $\Gamma$'s, which describe the coupling between the molecule and the contacts~\cite{Ryndyk2016}. When the contacts are metallic, we can take the wide-band limit since their density of states are flat. Thus, the $\Gamma$'s are simply given by a constant, which we take as the energy unit: $\Gamma^L_{\mu 1 \sigma, \mu 1 \sigma}=\Gamma^R_{\nu N s, \nu N s}=\Gamma_0=1$ ($\mu,\nu = A,B$, and $\sigma, s=\uparrow,\downarrow$). The sum over strands will provide us with the electron transmission per spin 
\begin{eqnarray}
 \mathcal{T}_{\sigma, s}(\omega) &=& \sum_{\mu\nu} \mathcal{T}_{\mu \sigma,\nu s}(\omega) \ .
\end{eqnarray}
%
We are now in a position to assess the spin polarization generated in the molecular transport by means of the following energy-dependent quantity
\begin{eqnarray}\label{Eq:SpinPolarization}
\mathcal{P}(\omega) = \frac{\mathcal{T}_{\uparrow,\uparrow}(\omega)+\mathcal{T}_{\uparrow,\downarrow}(\omega)-\mathcal{T}_{\downarrow,\uparrow}(\omega)-\mathcal{T}_{\downarrow,\downarrow}(\omega)}{\mathcal{T}_{\uparrow,\uparrow}(\omega)+\mathcal{T}_{\uparrow,\downarrow}(\omega)+\mathcal{T}_{\downarrow,\uparrow}(\omega)+\mathcal{T}_{\downarrow,\downarrow}(\omega)} \ ,
\end{eqnarray}
which will be referred to as spin polarization in what follows.

The transmission completely determines the current for small applied voltages at very low temperature. This description is close to the experiments of Ref.~\cite{SMLL:SMLL201602519} that were performed in the linear response regime with a small applied bias (50 mV). In such a case, a spin polarization of nearly $50\%$ from the linear conductance of the junction was obtained. Thus, we claim that the transmission function, which is proportional to the conductance at very low temperature, may still be a good reference point for the investigation of spin polarization. In different experimental setups, additional non-equilibrium effects play a role but these are beyond the scope of this work, although they are expected to magnify the CISS effect.

\subsection{Analysis of symmetries}
\label{Symm}

\begin{figure}
 \centering
 \includegraphics[scale=0.6]{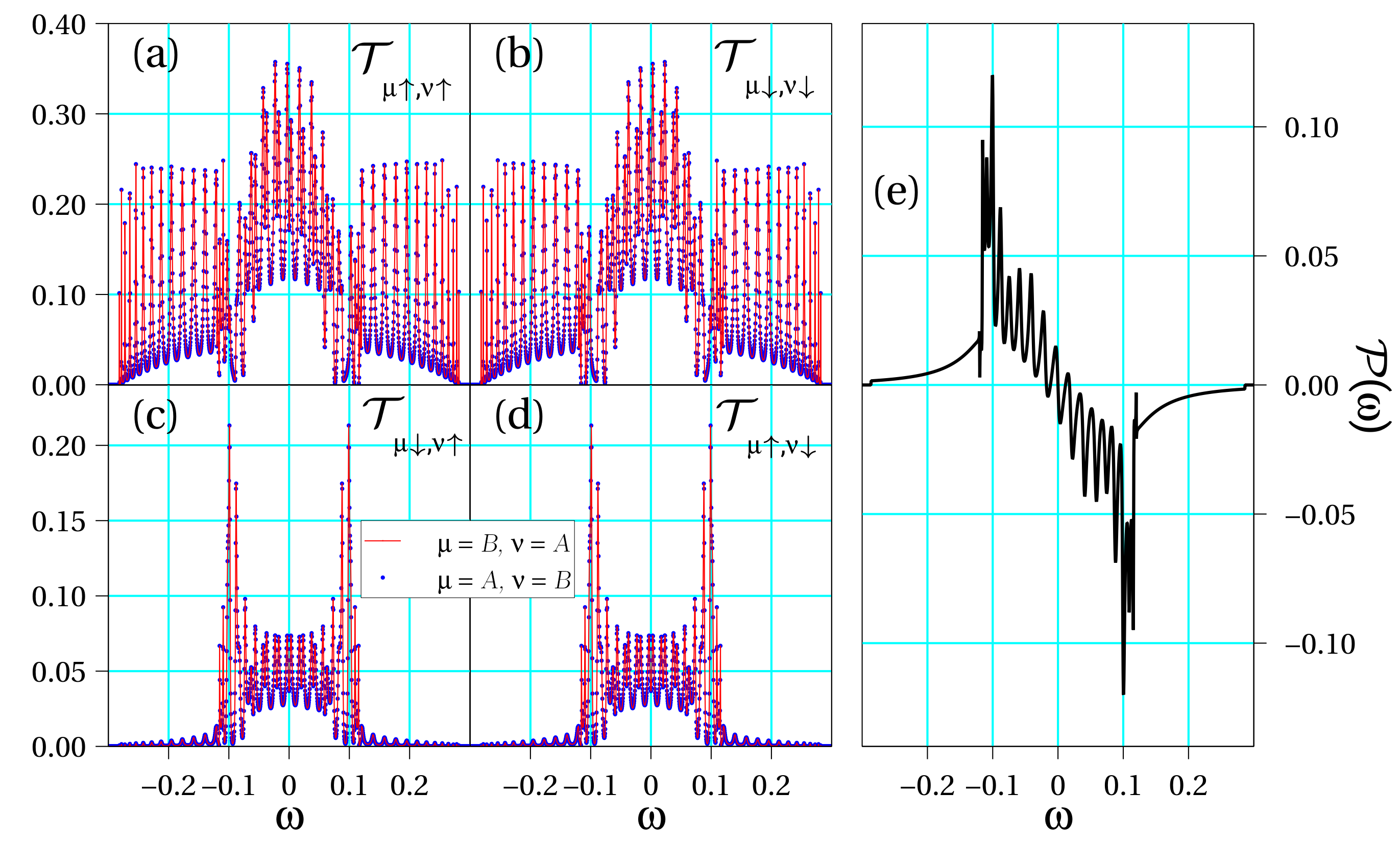}
 \caption{Transport coefficients obtained for a symmetric molecule: $\varepsilon_A=\varepsilon_B=0.0$, $t_A=t_B=0.1$, $\lambda_A = \lambda_B = 0.01$, $\alpha_c = -0.08$, $\theta = 0.66$, $N_{st}=10$, $N_{tr}=3$ and $\Gamma_0 = 1.0$. Transmission coefficients (a) $\mathcal{T}_{\mu\uparrow,\nu\uparrow}(\omega)$, (b) $\mathcal{T}_{\mu\downarrow,\nu\downarrow}(\omega)$, (c) $\mathcal{T}_{\mu\downarrow,\nu\uparrow}(\omega)$, (d) $\mathcal{T}_{\mu\uparrow,\nu\downarrow}(\omega)$ and  (e) spin polarization $P(\omega)$ as a function of energy.}
 \label{fig:2}
\end{figure}

\begin{figure}
 \centering
 \includegraphics[scale=0.6]{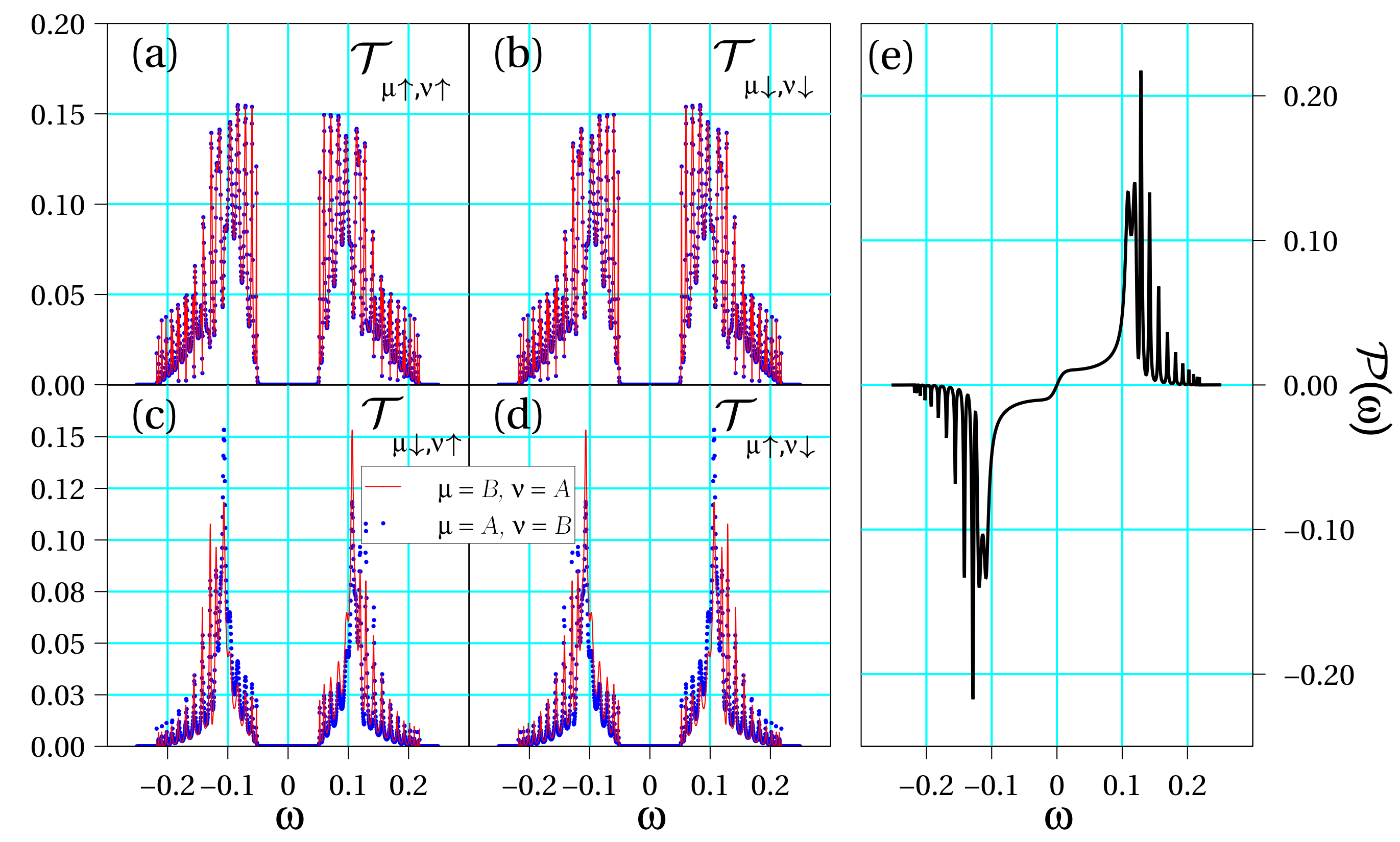}
 \caption{Transport coefficients obtained for a quasi-symmetric molecules with the same parameters as in Fig.~\ref{fig:2} but $t_A = -t_B = 0.1$ and $\lambda_A = \lambda_B = 0.01$. Transmission coefficients (a) $\mathcal{T}_{\mu\uparrow,\nu\uparrow}(\omega)$, (b) $\mathcal{T}_{\mu\downarrow,\nu\downarrow}(\omega)$, (c) $\mathcal{T}_{\mu\downarrow,\nu\uparrow}(\omega)$, (d) $\mathcal{T}_{\mu\uparrow,\nu\downarrow}(\omega)$ and (e) spin polarization $P(\omega)$ as a function of energy.}
 \label{fig:3}
\end{figure}

\begin{figure}
 \centering
 \includegraphics[scale=0.6]{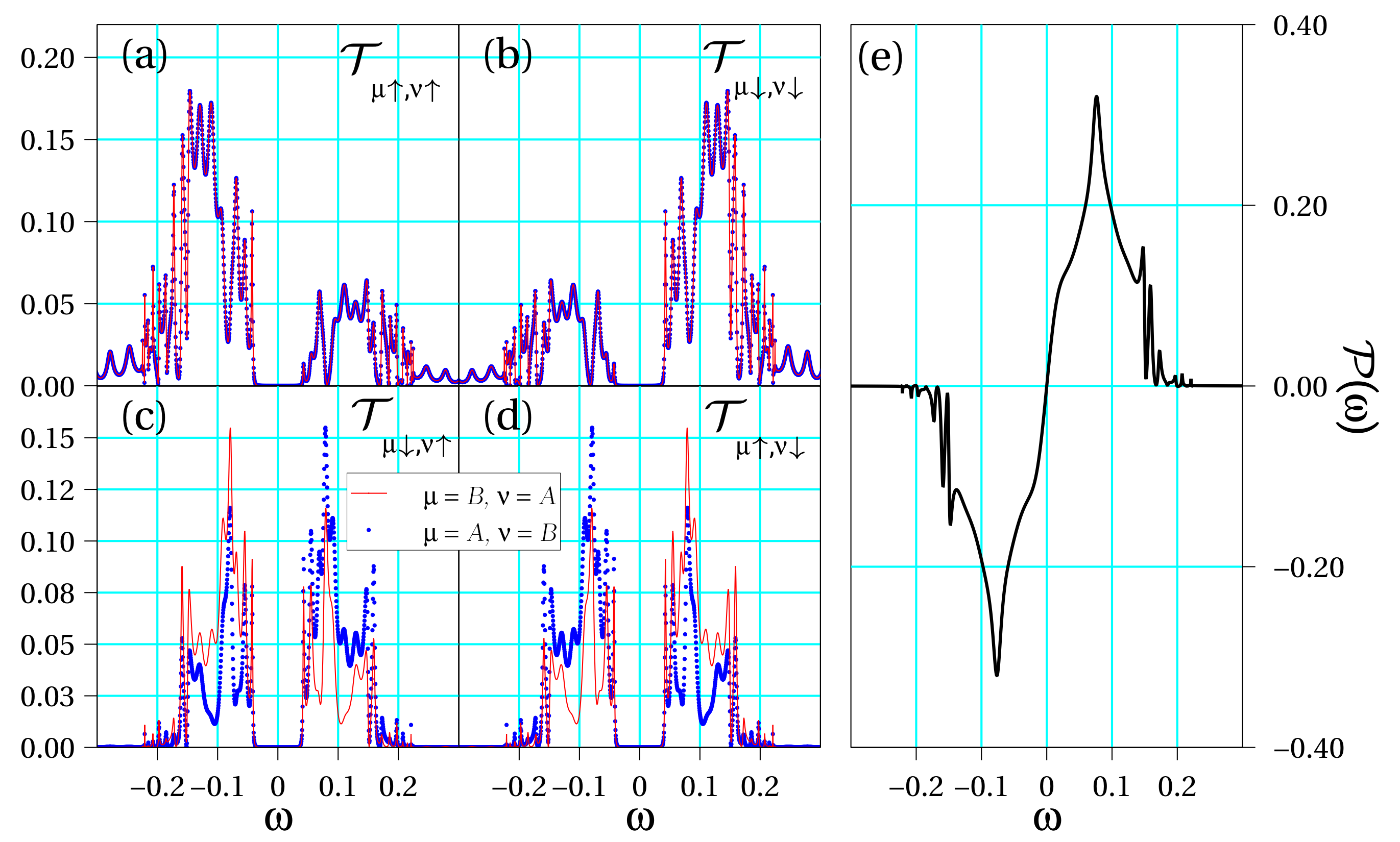}
 \caption{Transport coefficients obtained for an asymmetric molecule with the same parameters as in Fig.~\ref{fig:2} but $t_A = -2t_B = 0.2$ and $2\lambda_A = \lambda_B = 0.02$. Transmission coefficients (a) $\mathcal{T}_{\mu\uparrow,\nu\uparrow}(\omega)$, (b) $\mathcal{T}_{\mu\downarrow,\nu\downarrow}(\omega)$, (c) $\mathcal{T}_{\mu\downarrow,\nu\uparrow}(\omega)$, (d) $\mathcal{T}_{\mu\uparrow,\nu\downarrow}(\omega)$ and (e) spin polarization $P(\omega)$ as a function of energy.}
 \label{fig:4}
\end{figure}

Let us analyze the results derived from equations~\eqref{Eq:Transmission} and~\eqref{Eq:SpinPolarization} for the three representative cases considered in Sec.~2.2, namely, symmetric, quasi-symmetric and asymmetric configuration. Since the calculations are fully numerical, we take into account the site dependence of the SOC potential. In figures~\ref{fig:2}--\ref{fig:4} we show all interstrand spin-flip components of the transmission in addition to the spin polarization for each set of parameters. For concreteness, hereafter we will consider a DNA molecule of $N_{t}=3$ turns in its B-form, which has $N_{b}=10$ bases in a complete turn. Hence, $\varphi=2\pi/10$. As a result, our simulations will contain $N=N_{b}N_{t}=30$ sites. Ab-initio calculations calculations suggest values of the intrastrand hopping roughly in the range of  $30-80\,$meV (see, e.g., Ref.~\cite{Kubar2008}) in DNA and we take $50\,$meV as a typical value. Assuming a lead-molecule coupling of the order of $\Gamma_0 \sim 250\,$meV yields $t_\mu/\Gamma_0\sim 0.2$. As to the SOC, we will take $\lambda_\mu \sim 5\,$meV, hence $\lambda_\mu/\Gamma_0 \sim 0.02$.

Figure~\ref{fig:2} shows the curves $\mathcal{T}_{\mu \sigma,\nu s}(\omega)$ in different panels for the symmetric situation, namely, when the physical parameters of the two DNA strands are the same. Later, as an intermediate case, figure~\ref{fig:3} shows the same magnitudes when a small asymmetry is introduced in the double-stranded molecular hopping coefficients, $t_A=-t_B=0.1$ (recall that $\Gamma_0$ is set as the unit of energy), but the SOC strengths are equal. Finally, figure~\ref{fig:4} summarizes the same physical magnitudes when both DNA strands are clearly asymmetrical with regard to both their different electronic hoppings, $t_A=-2t_B=0.2$, and SOC intensities, $2\lambda_A=\lambda_B=0.02$. 

In all considered cases, we observe that there exist contributions for each set of spins. However, it is most important to mention that the following symmetry condition for spin-flip processes arise from our calculations:
\begin{eqnarray}\label{Eq:Spinsym}
 \mathcal{T}_{\mu \uparrow,\nu \downarrow}(\omega)=\mathcal{T}_{\nu \downarrow ,\mu \uparrow}(\omega) \ .
\end{eqnarray}
This originates from the fact that the transmission of a Rashba
conductor is, quite generally, independent of the magnetization direction, a statement that has been demonstrated for one-dimensional (1D)~\cite{Molenkamp01} and quasi-1D systems~\cite{Sanchez06}. Furthermore, for spin-conserving transmission contributions, we observe additional symmetries:
\begin{eqnarray}\label{Eq:helixsym1}
\mathcal{T}_{\mu \sigma,\nu \sigma}(\omega)&=&\mathcal{T}_{\nu \sigma,\mu \sigma}(\omega)\ ,\nonumber \\
\mathcal{T}_{\mu \uparrow,\nu \uparrow}(\omega)&=&\mathcal{T}_{\mu \downarrow,\nu \downarrow}(-\omega) \ .
\label{Eq:helixsym2}
\end{eqnarray}
Moreover, the following symmetry conditions for intrastrand contributions also hold (not explicitly shown in the figures for brevity):
\begin{eqnarray}
\mathcal{T}_{\mu \uparrow,\mu \downarrow}(\omega)&=&\mathcal{T}_{\mu \downarrow,\mu \uparrow}(\omega)\ ,\nonumber \\
\mathcal{T}_{\mu \uparrow,\mu \uparrow}(\omega)&=&\mathcal{T}_{\mu \downarrow,\mu \downarrow}(-\omega) \ .
\label{Eq:helixsym3}
\end{eqnarray}
The latter conditions lead to the full cancellation of the spin-flip processes such that the components of the transmission which play a role in the spin polarization are $\mathcal{T}_{\uparrow \mu,\uparrow \nu}(\omega)$ and $\mathcal{T}_{\downarrow \mu,\downarrow \nu}(\omega)$, including the case $\mu=\nu$. Therefore, the resulting spin polarization reads
\begin{eqnarray}
 \mathcal{P}(\omega) = \frac{\sum_{\mu\nu}\left[\mathcal{T}_{\mu \uparrow,\nu \uparrow}(\omega) - \mathcal{T}_{\mu \downarrow,\nu \downarrow}(\omega)\right]}{\mathcal{T}_{\uparrow,\uparrow}(\omega)+\mathcal{T}_{\uparrow,\downarrow}(\omega)+\mathcal{T}_{\downarrow,\uparrow}(\omega)+\mathcal{T}_{\downarrow,\downarrow}(\omega)}\ .
\end{eqnarray}
In fact, the symmetries shown in equations~\eqref{Eq:helixsym2} and \eqref{Eq:helixsym3} imply an energy symmetry in the polarization, which reads
\begin{eqnarray}\label{Eq:Polarisym}
 \mathcal{P}(\omega) = -\mathcal{P}(-\omega)\ .
\end{eqnarray}

\section{Discussion}

In this work we revisit one of the most relevant model Hamiltonians widely used to describe spin-dependent transport in chiral molecular systems~\cite{Guo12,Gutierrez13}. In particular, we focus on a description that accounts for a double-stranded helical molecule to mimic DNA molecules in its most common structure, namely, B-form DNA. It is worth noticing that in our study we keep the physical parameters within reasonable values accepted for DNA $\lambda_{\mu} < |\alc| \sim |t_{\mu}|$. We demonstrate that a sizable spin polarization arises in chiral systems with no need of including dephasing effects if the terminal connections are properly simulated~\cite{Guo14a}. In such scenario the asymmetry between the two DNA strands turns out to be a trigger for the spin polarization to arise. Here, we consider three representative cases to analyse this issue. Our reference system is referred to as symmetrical case, where the two strands are equal. We also consider two other situations where this condition is relaxed. First, in view of some first-principle calculations~\cite{Artacho03}, we introduce a small asymmetry in the two strands by considering their electronic hoppings with opposite signs~\cite{Diaz07}. Lastly, we obtain spin polarization in a very asymmetrical situation where the two DNA strands are described by different electronic and spin-orbit couplings. The latter being a limiting case for which it was proven that the spin polarization is clearly enhanced~\cite{Gutierrez13,Guo14a}. In addition, one of the novel interests of the present study is the analysis of the symmetries that arise in the transmission coefficients involved in the spin transport along chiral molecular systems.

Let us now summarize the main conclusions of our work. On the one hand, from figures~\ref{fig:2} and~\ref{fig:3} it is clear that even a tiny seed of asymmetry between the two DNA strands lead to a clear enhancement of the spin polarization in the molecule. In particular, just a change of the electronic hopping sign doubles the magnitude of the spin polarization although the energy window within which this enhancement is relevant is still small. However, when the asymmetry between the two strands is further increased as in figure~\ref{fig:4}, we obtain an even higher spin polarization and, most importantly, it happens in a clearly wider energy window. From our simulations, not shown here for brevity, it is clear that the opposite signs of the electronic hoppings turns out to be a crucial ingredient to get this effect. This fact is related to the band structures shown in figure~\ref{Fig:0}, where anti-crossing points arise only if the hoppings parameters have opposite signs. In previous studies~\cite{Gutierrez13,doi:10.1021/acs.jpcc.9b07764} it was shown that the non-vanishing of spin polarization within a two-strand helical model as the one used in the present study, is  related to the impossibility of fully removing the SOC from the model via a unitary transformation. In particular, it turns out that asymmetries in the values of the electronic couplings and/or in the strength of the SOC can lead to an increased polarization, a result nicely confirmed in the current investigation. We remark, however, that additionally both strands need to be connected to left and right electrodes in order to get a non-zero polarization.

Regarding the transmission coefficient symmetries, figures~\ref{fig:2}--\ref{fig:4} clearly establish the symmetries summarized in equations~(\ref{Eq:helixsym1}) and (\ref{Eq:helixsym3}). These symmetries will be valid as long as the  energies of the molecular orbitals of the two strands are equal or symmetric around zero. If this is not the case, then the conditions should be slightly modified. This analysis allows us to demonstrate that the processes responsible for the spin polarization in chiral molecules are those that conserve the spin state. Indeed, it seems plausible that even in other possible configurations, the spin-flip contributions will be less significant. Our results based on and effective model are in perfect agreement with other recent statements based on {\em ab-initio}  calculations~\cite{doi:10.1021/acs.jpclett.8b02360} and rate equations for the analysis of photoemission experiments~\cite{C8CP07257D}.

\acknowledgments

We thank Matthias Geyer and Arezoo Dianat for fruitful discussions. This research was funded by the Spanish Ministry of Science, Innovation and Universities, grant numbers MAT2016-75955 and MAT2017-82639. R. G. and G.C. acknowledge financial support from the Volkswagen Stiftung (grant number 88366). This work was partly supported by the German Research Foundation (DFG) within the Cluster of Excellence \textquotedblleft Center for Advancing Electronics Dresden\textquotedblright. M. S. acknowledges financial support by the Deutsche Forschungsgemeinschaft (DFG, German Research Foundation), Project-id No. 258499086—SFB1170 “ToCoTronics”, the Free State of Bavaria (Elitenetzwerk Bayern IDK “Topologische Isolatoren”) and the Würzburg-Dresden Cluster of Excellence on Complexity and Topology in Quantum Matter—ct.qmat(EXC 2147, grant number 39085490).

\bibliography{references}

\end{document}